\documentclass[reprint,twocolumn,showpacs,nofootinbib,notitlepage,footnotes]{revtex4-1}
\usepackage{enumerate}
\usepackage{graphicx}
\usepackage{latexsym,amsmath,amssymb,lmodern,float,url}
\usepackage{natbib}
\usepackage{color}
\usepackage{microtype}
\usepackage{slashed}
\usepackage{comment}
\usepackage{ulem}
\usepackage[colorlinks=true,backref=false, linktocpage=true,
citecolor=blue,urlcolor=blue,linkcolor=blue,pdfpagemode=UseOutlines]{hyperref}

\hypersetup{%
  bookmarksnumbered=true,
  pdftitle = {},
  pdfsubject = {},
  pdfauthor = {},
  pdfkeywords = {}
}

\DeclareMathOperator{\Tr}{Tr}

\begin{document}

\title{Pure gauge theories and spatial periodicity}
\author{Thomas D. Cohen}
\email{cohen@umd.edu}
\affiliation{Department of Physics, University of Maryland, College Park, Maryland 20742, USA}

\date{\today}

\begin{abstract}
Properties of pure gauge theories in thermal equilibrium as calculated via standard functional integral treatments are mathematically identical to  ground state properties of a theory with spatially-periodic boundary conditions imposed on the gauge fields.  Such a theory has states that have no analog in a  theory in which only physical observables  associated with gauge-invariant operators are required to be periodic, rather than the gauge fields themselves;  these states are in topological sectors that do not exist in the unconstrained theory. The topology arises because the boundary conditions in the functional integral are gauge invariant on a cylinder but not in the unconstrained theory. 
\end{abstract}

\maketitle

\section{Introduction}

Confinement has been a key theoretical issue ever since QCD emerged in the early 1970s\cite{Fritzsch:1973pi}.     However,  what is even meant by  ``confinement'' and ``deconfinement''  is not totally clear\cite{Greensite:2011zz}.  Some issues appear to be framed more sharply in pure gauge (Yang-Mills) theories rather than in QCD itself as this allows a focus on quantities associated with the Polyakov loop\cite{Polyakov:1978vu}. In the high temperature phase of Yang-Mills, the Polyakov loop appears to acquire an expectation value, implying a spontaneous breaking of center symmetry and which is taken to mean deconfinement.  Reviews  of the Polyakov loop expectation value as a signature of confinement are found  in ref.~\cite{Fukushima:2017csk,Greensite:2011zz}.  It is not totally straightforward as a signature: extracting  an expectation value of the Polyakov loop directly from lattice studies was recognized to be subtle from the earliest days \cite{McLerran:1981pb}.  For example, there are serious renormalization issues that naturally  require a calculation of the correlator of  a  Polyakov loop  and its conjugate to resolve\cite{2002}.    Indeed, nearly three decades ago  Smilga \cite{Smilga:1993vb}  raised fundamental questions about whether a single Polyakov loop can acquire a expectation value.
 
The  use of the string tension extracted from the correlator of a Polyakov loop and its conjugate\cite{McLerran:1981pb,Greensite:2011zz} as a signature of confinement evades these subtleties. The correlator has been interpreted as the exponential of the free energy of an arbitrarily heavy quark and anti-quark injected into the system and held a fixed distance apart\cite{Polyakov:1978vu,McLerran:1981pb,Greensite:2011zz}.  If the free energy increases linearly, the  string tension  is non-zero as expected in a confined phase.  A vanishing string tension then becomes a signature of  deconfinement. Order parameters that vanish in one phase and not in another are often associated with a symmetry and it's spontaneous breaking.   Since Polyakov's original paper \cite{Polyakov:1978vu}, center symmetry has been identified as the relevant symmetry.   Lattice studies beginning four decades ago \cite{McLerran:1981pb} have shown behavior consistent with a non-vanishing string tension in a low temperature phase and a vanishing string tension in a high temperature phase.

This interpretation is not entirely straightforward: the string tension is based on Polyakov loops, which have a clear meaning in functional integral treatments in Euclidean space, but whose meaning in Minkowski space remains rather obscure.  This is problematic since the dynamics of physical systems plays out in Minkowski space. This issue has been of limited practical significance in numerical studies to date: despite numerous attempts to tame it\cite{alexandru2020complex,Aarts:2015tyj,deForcrand:2009zkb,2000Karsch}, the sign problem continues to plague computations of many quantities in QCD, including real-time dynamics.  However, recent developments in quantum computing and its possible application to gauge theories\cite{Stryker:2018efp,Alexandru:2019nsa,Lamm:2019uyc,Lamm:2020jwv,deJong:2020tvx,Yamamoto:2020eqi,Gustafson:2020yfe,Ji:2020kjk,Davoudi:2020yln,Ciavarella:2021nmj,Cohen:2021imf} have raised the prospect of computations of real-time dynamics for gauge theories. 

Thus, it has become increasingly important to fully explore the physical significance of the Polyakov loop, center symmetry and the finite temperature  string  tension in order to understand their possible role in real-time dynamics.

\section{Pure gauge theory on a cylinder}

To gain insight into  the Polyakov loop, center transformations and the string tension, it is useful to consider quantities that are mathematically identical but which occur in a very different---and potentially more transparent---physical context.  In Euclidean space-time, time is equivalent to any spatial direction: a functional integral whose integrand is constrained to be periodic in a spatial direction (for specificity $z$) is identical to one that is periodic in Euclidean time (and used to describe thermal systems)\cite{2012R,2016R1,2016R2}.   While the basic idea of exploiting the formal identity of  compactification in space rather than Euclidean time for gauge theories is not new, the present paper elucidates the role of topology. Importantly, it  demonstrates that a theory compactified to a cylinder has states that have no analog in a  theory in which only physical observables---those associated with gauge-invariant operators in the uncompactified theory---are required to be periodic, rather than the gauge fields themselves.   These new states are in topological sectors that do not exist in the unconstrained theory.  The spatial analog of the Polyakov loop will be shown to bring the compactified theory from one topological sector to another.

This paper considers the pure gauge theory in which  $x$, $y$ and Euclidean time $\tau$ are unconstrained but where $z$ is periodic with periodicity $\beta=1/T$  (where $T$ is the temperature of the associated thermal system):
\begin{equation}
A_\mu(\vec{x}, \tau) = A_\mu(\vec{x} +\beta \hat{z}, \tau)\; . \label{Eq:spatialBC}
\end{equation}
This system is gauge invariant on a physical space with the topology of a cylinder;  the point $\vec{x} +\beta \hat{z}$ is identical to the point $\vec{x}$.  This paper considers pure gauge theory for SU(3), but {\it mutatis mutandis}, the analysis goes through for other gauge groups.

The spatial analog of the usual  Polyakov loop is 
\begin{equation}
\begin{split}
L_S(\vec{x,y,\tau}; \beta) &\equiv \Tr \left [ \overleftrightarrow{L}_s(\vec{x,y,\tau}; \beta)  \right] \\
 \overleftrightarrow{L_s}(x,y,\tau; \beta) &  \equiv P \left (\exp \left (i \int_{0}^{\beta} \, d z  A_z(x,y,z,\tau) \right ) \right ) 
\end{split}\label{Eq:PolyS}
\end{equation}
where $P$ indicates  path-ordered; the subscript $S$ indicates that it is spatial.  On the cylinder, $L_S$  is gauge invariant.  $C_{L_S^* L_S} (x,y,\tau; \beta)$,  the correlator of a spatial Polyakov loop and its conjugate separated by Euclidean time, $\tau$, is given by
\begin{equation} 
\begin{split}
&C_{L_S^* L_S} (x,y,\tau; \beta) =  \\
&\langle {\rm vac}(\beta) |  L_S^*(x,y,\tau;\beta) L_s(x,y, 0, \beta) |{\rm vac}(\beta) \rangle \; ,
\end{split} \end{equation}
$|{\rm vac}(\beta) \rangle$ is the vacuum for the  system on a cylinder with circumference $\beta$. 

Similarly, there is a spatial version of a center transformation, ${\cal C}_S$.  Locally ${\cal C}_S$ acts like a gauge transformation parameterized  by a function $\Omega(x,y,z,\tau) \in {\rm SU}(3)$.  However gauge transformations must be continuous on the cylinder, which requires periodicity; $\Omega(x,y,z,\tau)=\Omega(x,y,z+\beta,\tau)$.  In contrast, ${\cal C}_S$ has $\Omega(x,y,z,\tau)=\Omega(x,y,z+\beta,\tau) \exp \left (  \frac{2 \pi i}{3} \right )$,  making $\Omega(x,y,z,\tau)$ discontinuous on the cylinder.  However, it preserves the periodicity of the gauge field so that $A_\mu$ remains continuous.    

$L_S$, $C_{L_S^* L_S}$ and ${\cal C}_S$ are constructed so that when computed via functional integrals on a spatial cylinder, they behave identically with their temporal analogs---which have standard interpretation in terms of thermal physics.  Thus, the string tension for the thermal system at temperature $T=1/\beta$ can be read off directly from $C_{L_S^* L_S} (x,y,\tau; \beta)$:
\begin{equation}
 \sigma = \lim_{\tau \rightarrow \infty} \frac{- \log \left (C_{L_S^* L_S} (x,y,\tau; \beta)) \right )}{\beta  \tau} \; . \label{Eq:sigma}
\end{equation}

While these quantities are mathematically identical to the analogous thermal ones, their physical meaning are quite different.  The correlator of two Euclidean-time separated spatial Polyakov loops has a  simple interpretation
\begin{equation}
\begin{split}
&C_{L_S^* L_S} (x,y,\tau; \beta) =\\ 
&\langle {\rm vac}(\beta) | e^{\beta H}  L_S^*(x,y,0;\beta) e^{-\beta_H} L_s(x,y, 0; \beta) |{\rm vac}(\beta) \rangle\\ 
&= \sum_{k \in {\rm phys}} \exp(- \beta E_i) \, |\langle \psi_k |L_s(x,y, 0; \beta)|\rm vac (\beta) \rangle|^2 ,
\end{split} \label{Eq:corr} 
\end{equation}
where $E_i$ is the energy relative to the vacuum.   The energy of the  lowest-lying excited state connected to the vacuum via $L_S$ will be denoted $E_1$. A thermal theory with $T=1/\beta$ has
\begin{equation}
\sigma(T) = T E_1 \; .
\label{Eq:sigmaT}\end{equation}

${\cal C}_S$ differs from an ordinary gauge globally, but locally it behaves as a gauge transformation.  Thus any locally gauge-invariant quantities, $Q(\vec{x},t)$, is unaffected: $ {\cal C}_S^\dagger \, Q(\vec{x},t) \,  {\cal C}_S = Q(\vec{x},t)$.  The stress-energy tensor is  gauge-invariant and local,  so $[T_{00}, {\cal C}_S]=0$, implying
\begin{equation}
[H,  {\cal C}_S] =0 \; ,
\end{equation}
since $H$ is  the spatial integral of $T_{00}$.  Thus, energy eigenstates can be divided into sectors labeled by $n=0,1,2$, where $ e^{ i \frac{2 n \pi}{3}}$ are the three eigenvalues of  ${\cal C}_S$.

Spatial Polyakov loops are not invariant under spatial center transformations:
\begin{equation}
\begin{split}
&{\cal C}_S^\dagger L_s(x,y,\tau;\beta)  {\cal C}_S = \exp \left ( i \frac{2 \pi }{3} \right ) L_s(x,y,\tau;\beta) \;,\\ 
&{\rm so \; that\; if} \; {\cal C}_S |\psi\rangle=  e^{ i \frac{2 \pi }{3} } |\psi\rangle\;  \; {\rm then}\\
&\!\!\ {\cal C}_S\left ( L_s(x,y,\tau;\beta)  |\psi\rangle \right ) =   e^{ i \frac{2 \pi (n+1)}{3} }  ( L_s(x,y,\tau;\beta)  |\psi\rangle  ) \; .
\end{split}
\label{Eq:actionCs}\end{equation}
When a spatial Polyakov loop acts on a state,  it produces a state in a  different sector from the original. For example,  if  $|\psi\rangle$ is in the  $n=0$ sector, then   $( L_s(x,y,\tau;\beta)  |\psi\rangle  )$ has $n=1$.    

The vacuum state in the large $\beta$ (``confined'') phase of the theory on the cylinder, must have $n=0$.    Under  parity in three spatial dimensions  ${\cal C}_S \rightarrow {\cal C}_S^\dagger$.  This implies that  for every energy eigenstate with $n=1$  there  is a degenerate one with $n=2$.  If the vacuum is not in the $n=0$ sector, then it must be (at least) doubly degenerate and spontaneously break parity.  At infinite $\beta$, the theory is just ordinary Yang-Mills, which is known not to break parity spontaneously\cite{Vafa:1984xg}.  The theory will remain in a phase with unbroken parity requiring that the vacuum has $n=0$, until a phase transition is reached.  

From Eq.~(\ref{Eq:actionCs}), the string tension (in the thermal context)  is given by the temperature times the energy of the lightest state in the system on the cylinder carrying the quantum number $n=1$. 

As the circumference of the cylinder decreases, the system undergoes a first order phase transition to a phase with a string tension of zero in which $L_S$ operating on the vacuum either annihilates it or produces a state degenerate with it.  The first possibility can be eliminated: $C_{L_S^* L_S} (x,y,\tau; \beta)$, the correlator of Polyakov loop, is nonzero regardless of the phase as its short distance behavior is fixed\cite{2002}; the vacuum must be (at least) three-fold degenerate and the system spontaneously breaks the spatial center symmetry\footnote{One might worry that the spontaneous breaking of center symmetry implies the spontaneous breaking of parity and runs afoul of the argument by Vafa and Witten\cite{Vafa:1984xg} that parity cannot spontaneously break.  That is not a concern here.   The Vafa-Witten argument  applies to the vacuum of Lorentz-invariant theories in 3+1 dimensions as it relies on the very limited set of parity-violating operators allowed in these circumstances\cite{Cohen:2001hf}.  In other situations, parity is known to break spontaneously such as pion condensation of QCD with an isospin chemical potential \cite{Son:2000xc,Son:2000by}  }.  

 It is important to note that while the theory breaks a symmetry spontaneously, there are no Goldstone modes as the symmetry is discrete.
 
 \section{Periodicity without topology}

 The theory  on the cylinder is connected in an interesting way to the theory in infinite  flat space but with all physical quantities constrained to act periodically: ${\cal O}(\vec{x},\tau)={\cal O}(\vec{x}+\beta \hat{z},\tau )$ holds for every allowed physical operator,  as it does on the cylinder.   Normalization simplifies with  a large finite system  with periodic boundary conditions in $z$, with periodicity $N_z \beta$ where $N_z$ is a large positive integer ( the system contains $N_z$ periods of length $\beta$); the limit $N_z \rightarrow \infty$ can be taken at the end.   This will be denoted the ``periodically-constrained large system" (PCLS).  The ground state of the PCLS is labeled $| \psi_0; N_z ,\beta \rangle$.  The PCLS has the same Hamiltonian density as the cylinder and is designed  for the physics to have the same periodicity as the cylinder, so it is intuitively plausible that the two theories are equivalent in some sense---but the sense of equivalence is subtle.

To  specify the ground state of the periodically-constrained  large system, it is useful to define the spatial  average of a generic gauge-invariant operator ${\cal O}$ over $N_z$ periods:
\begin{equation}
\overline{{\cal O}} \equiv \frac{\sum_{k=0}^{N_z-1} e^{-i  k P_z  \beta} {\cal O} e^{i k P_z \beta}}{N_z}  \label{Eq:OpAv}
\end{equation}
where $P_z$ is the momentum operator.  $\overline{{\cal O}}$ is the projection of  ${\cal O}$ onto its periodic part.  Operators  on the cylinder satisfy  $\overline{{\cal O}} =  {\cal O }$.  Gauge-invariant operators in the PCLS that do not satisfy   $\overline{{\cal O}} =  {\cal O }$ have no analog on the cylinder, while   operators  that satisfy $\overline{{\cal O}} =  {\cal O }$ do.  Indeed, they are more than mere analogs: written as functions of the gluon fields and their conjugate momenta, they are formally identical---with the  caveat that on the cylinder points  separated in $z$ by $\beta$ are regarded as identical, while in the PCLS they are physically equivalent but distinct.    For clarity, the operators defined on the cylinder will have a superscript ${\rm cyl}$.
  
$| \psi; N_z ,\beta \rangle$, is a generic state of the periodically-constrained large system, can contain no physical correlations incommensurate with the periodicity of the cylinder.  Thus, 
\begin{equation}
\langle \psi; N_z ,\beta | {\cal O} -\overline{\cal O} | \psi; N_z ,\beta \rangle = 0 \; . \label{Eq:ComCorr}
\end{equation}
for all gauge-invariant operators ${\cal O}$ in the unconstrained theory on the large space.  The state  $ |\psi_0; N_z ,\beta \rangle$, the ground state of the PCLS is the lowest energy that satisfies Eq.~(\ref{Eq:ComCorr}) for all gauge invariant operators; it is analogous to the vacuum state on the cylinder.  

It should be clear that all $m-$point correlation functions of gauge-invariant operators  that are periodic ({\it i.e} satisfying  $\overline{{\cal O}} =  {\cal O }$) on the larger system must be identical to their analogs on the cylinder;  
\begin{equation}
\begin{split}
&  \forall  \;   {\cal O}_1,  {\cal O}_2 , {\cal O}_3 \ldots  \; \; {\rm satisfying}\\
&  \overline{{\cal O}_1} =  {\cal O}_1, \;  \overline{{\cal O}_2} ={\cal O}_2 , \;   \overline{{\cal O}_3} ={\cal O}_3 \ldots \; \; , \\ 
& \langle  \psi_0; N_z, \beta |{\cal O}_1{\cal O}_2 \ldots {\cal O}_{m-1}   {\cal O}_m |  \psi_0; N_z \beta ,\rangle = \\
&\langle {\rm vac}(\beta) |{\cal O}_1^{\rm cyl}{\cal O}_2^{\rm cyl} \ldots {\cal O}_{m-1}^{\rm cyl}   {\cal O}_m^{\rm cyl} | {\rm vac}(\beta)\rangle
\end{split} \label{Eq:bigsys}
\end{equation}

The lesson of Eq.~(\ref{Eq:bigsys}) is that all information about ground state correlators for Yang-Mills theory constrained so that only physics periodic in $z$ is allowed can be extracted from the system restricted to the spatial cylinder.    However, the converse is not true.  Only states in the $n=0$ sector of the theory on the cylinder have analogs in the PCLS and only operators connecting states within the $n=0$ sector have analogs.

Consider ${\cal C}_S$:  on the cylinder this is  gauge-invariant; it acts as a global symmetry operator. However, in the pure gauge theory with larger extent in $z$, it is simply a gauge transformation and has no effect when acting on physical states (i.e. those satisfying the color Gauss law). In the physical subspace of the theory, this operator is the identity, its eigenvalues are unity and the states are all in the $n=0$ sector.
 
An operator in the theory on a cylinder that  changes an $n=0$  state to a state with a different $n$ must be gauge invariant to be meaningful.  However, it cannot be gauge invariant in the PCLS: ${\cal C}_S$ is a gauge transformation on that theory and commutes with all gauge-invariant operators.  For example, $L_S$ is gauge invariant under gauge transformations on the cylinder but  not gauge-invariant in the theory with large extent in $z$.    Thus, the correlator of Polyakov loops, $C_{L_S^* L_S} (x,y,\tau; \beta)$---the quantity that on the spatial cylinder allows the extraction of the string tension in thermal  systems---has no analog in the PCLS.

Restricting configurations of  the original theory via the the imposition of Eq.~(\ref{Eq:spatialBC}) causes the emergence of new sectors of the theory with physical states  not present in the initial theory.  These sectors reflect the topology of the cylinder which arises due to the imposition of gauge invariance:  If  the physics centered around the points $\vec{x} +\beta \hat{z}$ and $\vec{x}$  were  physically equivalent but distinct, the boundary condition of Eq.~(\ref{Eq:spatialBC}) would not be gauge invariant; it only becomes  gauge invariant if the point $\vec{x} +\beta \hat{z}$ is identified with $\vec{x}$ as on the cylinder.

\section{Functional Integrals}

It is instructive to formulate a functional integral treatment of the PCLS that introduces no new topological sectors but guarantees that gauge-invariant quantities are identical to those on a cylinder.  This is straightforward:  the functional integral is over all configurations  for which $A_{\mu}$ is gauge-equivalent (in the full space) to a configuration satisfying  Eq.~(\ref{Eq:spatialBC}).  Unlike the boundary condition of Eq.~(\ref{Eq:spatialBC}) itself, such a boundary condition is always gauge invariant and does not force the system onto a cylinder. 

{  In a continuum version of the theory the boundary condition imposed on the functional integral is simply that 
\begin{equation}
\begin{split}
A_\mu(\vec{x}, \tau) = &\Lambda(\vec{x} +\beta \hat{z}) A_\mu(\vec{x} +\beta \hat{z}, \tau)  \Lambda(\vec{x} +\beta \hat{z})^\dagger\\ & -\frac{i}{g} \left(\partial_\mu \Lambda(\vec{x} +\beta \hat{z}) \right)   \Lambda(\vec{x} +\beta \hat{z})^\dagger \; , \end{split} \label{Eq:spatialBCG}
\end{equation}
where $\Lambda(\vec{x},t) \in$ SU(3) is an arbitrary continuous and differentiable matrix-valued function of space and time that specifies a gauge transformation and does not have to be periodic.    The functional integral would be taken over all configurations consistent with Eq.~(\ref{Eq:spatialBC}).

Functional integrals for the continuum theory are rather formal objects.  To define the functional integral concretely one needs to regularize the theory and the most straightforward way to do this while respecting gauge invariance is via a lattice.  

One might worry about using a lattice to regulate the theory:  the issues under consideration here involve topology and the topology of a continuous space and of a set of discrete points connected by links are fundamentally different. However, the reason topological issues arose in this problem was as a result of the distinction between whether gauge fields themselves were periodic or whether the gauge fields could be aperiodic provided that all gauge invariant combinations of the gauge fields were periodic.  If the gauge fields themselves are periodic, a finite shift in the z-direction could be interpreted as bringing  the system back to the same physical point---as one would have with a generalized cylinder.  Conversely,if the gauge fields are not themselves periodic, the system cannot be on a generalized cylinder.  This distinction goes through analogously if one considers a lattice version of the theory (with periodicity of the link variables substituting for periodicity of the gauge fields).  Thus, one expects all of the analysis to through without change for a lattice version of the theory. Such a theory has well-defined functional integral representation and in as the lattice spacing approaches zero quantities should approach a well-defined limit associated with the continuum.   } 

Thus,  the most straightforward way to think about the functional integral is in a lattice-regularized version of the theory, with lattice spacing $a$.  In lattice units the period to be imposed is $n_\beta =\beta/a$ and the total number of lattice sites in the z direction is $N_z n_\beta$ (with periodic boundary conditions over the full space). 

Each link variable is specified by the position of the lattice site from which it starts and its direction;  a given link is specified $U_{n_x,n_y,n_z, n_\tau}^d$ where the direction, $d$, takes the values the $x,y,z$ and $\tau$ directions.  If a configuration  on the larger space satisfies
\begin{widetext}
\begin{equation}
\begin{split}
    & U_{n_x,n_y,n_z, n_\tau}^{x} = \Omega_{n_x+1,n_y,n_z+n_\beta,n_\tau} U_{n_x,n_y,n_z +n_\beta, n_\tau}^{x}  \Omega_{n_x,n_y,n_z+n_\beta,n_\tau}^\dagger \; ,\\
     & U_{n_x,n_y,n_z, n_\tau}^{y} = \Omega_{n_x,n_y+1,n_z+n_\beta,n_\tau} U_{n_x,n_y,n_z + n_\beta, n_\tau}^{y}  \Omega_{n_x,n_y,n_z+n_\beta,n_\tau}^\dagger \; ,\\
      & U_{n_x,n_y,n_z, n_\tau}^{z} = \Omega_{n_x,n_y,n_z+n_\beta +1,n_\tau} U_{n_x,n_y,n_z + n_\beta, n_\tau}^{z}  \Omega_{n_x,n_y,n_z+n_\beta,n_\tau}^\dagger \; ,\\
      & U_{n_x,n_y,n_z, n_\tau}^{\tau} = \Omega_{n_x,n_y,n_z+n_\beta,n_\tau+1} U_{n_x,n_y,n_z +n_\beta, n_\tau}^{\tau}  \Omega_{n_x,n_y,n_z+n_\beta,n_\tau}^\dagger \; ,\\
\label{Eq:percon}\end{split}\end{equation}\end{widetext}
(where the $\Omega$ are SU(3) matrices defined on the sites and define  gauge transformations), then in this configuration $\left(\overline{\cal O}-{\cal O}\right )=0$; as needed for states in the PCLS. 

In this construction neither the link matrices $U$  nor  $\Omega$ are constrained to have periodicity $\beta$.  However only the periodic part of any gauge-invariant quantity on the larger space is nonzero.  

To  specify a configuration uniquely, it is sufficient to specify a set of links, $\{U_1\}$ with elements $U_{n_x,n_y,n_z,n_\tau}^d$ for which $1 \le n_z < n_\beta$, and  $\{\Omega\}$, a set of gauge transformation matrices, with elements $\Omega_{{n_x,n_y,n_z,n_\tau}}$ with $n_\beta < n_z \le N_z n_\beta$.  Thus the functional integral expression for the expectation value of a gauge-invariant operator in the PCLS is given by
\begin{equation}
    \langle  {\cal O} \rangle  = \frac{\int d[U] \, \frac{ {\cal O}(\{U_1\},\{\Omega\} )}{N_Z} \, \,  \exp \left ( - \frac{S_{\rm YM}(\{U_1\},\{\Omega\}  )}{g^2 N_z} \right ) }{ \int d[U] \,  \exp \left ( - \frac{S_{\rm YM}(\{U_1\},\{\Omega\}  )}{g^2 N_z} \right ) } \label{Eq:FuncInt}
\end{equation}
where
\begin{widetext}
\begin{equation}
d[U] \equiv   \prod_{{\rm all} \, n_x} \, \prod_{{\rm all} \, n_y} \, \prod_{ {\rm all} \, n_\tau} \,  \prod_{n_z=1}^{n\beta} \, \prod_{d=x,y,z, \tau} \,  \prod_{n'_z=n \beta+1}^{N_z n\beta}  dU_{n_x,n_y,n_z,n_\tau}^d d\Omega_{{n_x,_y,n'_z,n_\tau}}  \; .
\end{equation}
\end{widetext}
The matrices are integrated with respect to the Haar measure, $S_{\rm YM}$ is the Yang-Mills action and $g$ is the coupling constant. ${\cal O}$ and the Yang-Mills action depend on the set of link variables at each site (and hence are full fixed by the matrices in the sets $\{U_1\}$ and $\{\Omega\}$).  ${\cal O}$ is divided by $N_Z$ to compensate for the fact that the functional integral sums over the operator $N_z$ times. In a normalized state in the Hilbert space this is compensated for by a factor of $N_z^{-\frac{1}{2}}$ for an overall factor of $1/N_z$; this is inserted by hand here in order to match that.   Similarly the Yang-Mills action is divided by $N_z$ since the action on the large space is $N_Z$ times the action in a single period while the integration variables correspond to a single period.  

By construction, $\langle \overline{\cal O}-{\cal O}\rangle =0$ as is required for the PCLS.  Moreover, ${\cal O}$  and $S_{\rm YM}$ are gauge invariant, so they only depend on $\{U_1\}$, the links in the first period.  The integrals over the gauge transformations yield identical overall constants in the numerator and denominator, which cancel. Thus, the result of a calculation for a  gauge invariant quantity calculated with this construction becomes identical to the result obtained with the boundary condition  Eq.~(\ref{Eq:spatialBC}), which requires that the system is on a cylinder.

However, if one focuses on quantities that are not gauge invariant in the large system, but are gauge invariant on the cylinder---such as correlators of Polyakov loops---this construction and the system on the cylinder differ radically.  Indeed, it is trivial to see that the correlator of the Polyakov loop vanishes identically with this construction.

\section{The string tension and the PCLS}

In the PCLS  the spatial Polyakov loop and its  correlators  are not gauge invariant and have no direct physical meaning.  In the functional integral of Eq.~(\ref{Eq:FuncInt}), it is easy to see that their expectation values vanish for any value of $\beta$.

While having no direct physical meaning in the PCLS, quantities such as the string tension (as extracted on the cylinder) can be used to infer properties of the PCLS indirectly; the value of the string tension as computed on the cylinder is correlated with the expectation value of operators in the analogous PCLS: expectation values of operators in the PCLS are identical to the expectation value of their analogs on the cylinder, which, in turn are correlated with the string tension.  

Thus a vanishing string tension on a cylinder can tell us whether the PCLS with the same $\beta$ is in the small-$\beta$ phase of theory.  In this sense, it acts as an order parameter---despite not being a direct  observable in the PCLS.
 
The distinction between direct and indirect  information about the PCLS can be significant.  Consider a sequence of states in which the physics is approximately, but not perfectly, periodic  and that approaches $ | \psi_0; N_z ,\beta \rangle$---the ground state of the PCLS---asymptotically. Suppose further that one wishes to study the expectation value of  operators  satisfying $\overline{{\cal O}} =  {\cal O }$ on states in the sequence. This is in some sense analogous to the study of a system described by a sequence of density matrices that  asymptotically approaches a density matrix describing a thermal distribution---as one might do to study a system thermalizing.  Operators in the PCLS that provide direct information about the system retain their physical meaning even when state  is only  approximately periodic.  Thus, expectation values of these observables are well defined for all states in sequence. This means that studies of the behavior of the expectation values of these operators are possible---at least in principle---and can yield information about the system as it approaches periodicity. In contrast, the string tension and other quantities providing only indirect information cannot be used to study behavior as the system approaches periodicity.   It is simply meaningless to ask how the string tension behaves until one reaches exact periodicity.

In summary, gauge-invariance and the imposition of spatial periodicity through the boundary condition Eq.~(\ref{Eq:spatialBC}) force the Yang-Mills system to be treated as being on a cylinder.  With this comes the emergence of topological sectors that were not present in the original theory, even when all physical properties of the original theory are constrained to have the same periodicity as the cylinder.  Since the Euclidean path integrals describing  thermal systems are identical to the those describing the spatial cylinder, the behavior on the cylinder may give insights as to the physical significance of the Polyakov loop and its correlators in thermal settings. This will be explored in a forthcoming publication, with an emphasis on the physical significance of the Polyakov loop, its correlators and their possible role in real-time dynamics.\\

\begin{acknowledgments}
Useful conversations with Lenya Glozman, Scott Lawrence, Owe Phillipsen and Yukari Yamauchi are gratefully acknowledged.   Masanori Hanada's suggestion to consider spatial periodicity helped prompt this work. This work was supported in part by the U.S. Department of Energy under Contract No.~DE-FG02-93ER-40762 and by the Maryland Center for Fundamental Physics .
\end{acknowledgments}

\bibliographystyle{apsrev4-1}
\bibliography{main}

\end{document}